\documentclass[preprint,showpacs,preprintnumbers,amsmath,amssymb]{revtex4}

\usepackage{graphicx}
\usepackage{dcolumn}
\usepackage{bm}

\begin{document}

\preprint{APS/123-QED}

\title{Mass independence and asymmetry of the reaction: Multi-fragmentation as an example\\} 

\author{Varinderjit Kaur}
\author{Suneel Kumar}%
 \email{suneel.kumar@thapar.edu}
\affiliation{
School of Physics and Materials Science, Thapar University Patiala-147004, Punjab (India)\\}
\date{\today}
\begin{abstract}
We present our recent results on the fragmentation by varying the mass asymmetry of the reaction between
0.2 and 0.7 at an incident energy of 250 MeV/nucleon. For the present study, the total mass of the system 
is kept constant (${A_{TOT}}$ = 152) and mass asymmetry of the reaction is defined by the asymmetry 
parameter 
(${{\eta}=  {\mid(A_T-A_P)}/{(A_T+A_P)}\mid}$). The measured distributions are shown as a function of
the total charge of all projectile fragments, ${Z_{bound}}$. We see an interesting outcome for rise and
fall in the production of intermediate mass fragments (IMFs) for large
asymmetric colliding nuclei. This trend, however, is completely missing for large asymmetric nuclei. 
Therefore, experiments are needed to verify this prediction.   
\end{abstract}
\maketitle
\baselineskip=25pt
\section{Introduction}
Heavy-ion collisions have always played a fascinating role in exploring various aspects of nuclear 
dynamics such as fusion-fission, multifragmentation and particle production. Multifragmentation, that is 
the emission of several intermediate mass fragments IMF's from a hot compound nucleus,
is a phenomenon observed in nuclear reactions over a wide incident energy range. There has been 
considerable progress during recent years in the experimental studies. Experimental evidence
for the statistical property of nuclear fragmentation has been given \cite{ogilvie,hubele,moretto} and 
various new quantities have been measured \cite{ogilvie,hubele,moretto,jain}. These quantities include 
the mean multiplicity of intermediate mass fragments 
(${<N_{IMF}>}$), the average charge of the largest fragment (${Z_{max}}$), the sum of all charges 
with ${Z \ge 2}$ etc. The quantity which is intimately related to the 
multifragmentation process is the multiplicity of intermediate mass fragments. Correlation between mean multiplicity of IMF's, ${<N_{IMF}>}$,
and the mass of the fragmenting system, whose measure is so called bound charge ${Z_{bound}}$ is an 
important aspect of multifragmentation that has been studied thoroughly by a number of groups 
\cite{hubele,jain,schuttauf}.
They, however, didnot take asymmetry of the system into account which is very important to study the
isospin effects \cite{zhang,liu01}.
The asymmetry of the reaction can be defined by the parameter ${{\eta}=  {\mid(A_T-A_P)}/{(A_T+A_P)\mid}}$;
 where ${A_T}$ and ${A_P}$ are the masses of target and projectile. The ${\eta}$ = 0 corresponds to the 
symmetric reactions, whereas, non-zero
value of ${\eta}$ define different asymmetry of the reaction. It is worth mentioning that the 
reaction dynamics in a symmetric reaction (${\eta}$ = 0) can be quite different compared to 
asymmetric reaction (${{\eta} \ne 0}$) \cite{donangelo}. This is due to the deposition of excitation 
energy in the form of compressional energy and thermal energy in symmetric and asymmetric reactions, 
respectively. The multifragmentation is studied many times in the literature 
\cite{zhang,liu01}. Unfortunately, very little study is available for the mass asymmetry of the reaction 
in terms of multifragmentation.\\     
In recent years, it has become possible to do exclusive measurements of multifragmentation process. This
has been done with streamer chamber detectors, electronic detectors, and ${4\pi}$ detectors. 
ALADiN \cite{hubele,jain,schuttauf} group has reported that the mean multiplicity of IMF's ${<N_{IMF}>}$ was found to be same 
for all targets ranging from Beryllium to Lead and for E/A ranging from 400 to 1000 MeV/nucleon.   
De Souza {\it et al.}, \cite{souza} observed a linear increase in the multifragmentation
of IMF's for central collisions with incident energies varying between 35 and 110 MeV/nucleon. In 2009,
Tsang {\it et al.}, \cite{tsang} reported a rise and fall in the production of IMF's. The maximal value of the IMF's
shifts from nearly central to peripheral collisions with the increase in the incident energy.\\  
Theoretically, multifragmentation can be studied by statistical \cite{statistical} as well as dynamical models
\cite{dynamical}. The universal property of multifragmentation has been quite satisfactorily described by the
statistical multifragmentation models \cite{statistical}. On the other hand, dynamical models are very 
useful for studying the reaction from the initial state to the final state where matter is fragmented
and cold. In this paper, we will adress the most interesting dependence of the multiplicity of intermediate mass fragments (IMF's). This multiplicity is estimated in terms of the ``bound'' charge value. We have used Isospin-dependent quantum molecular (IQMD) model to study 
the effect of asymmetry of colliding nuclei on the multifragmentation.\\

The isospin-dependent quantum molecular dynamics (IQMD)\cite{hartnack} model treats different 
charge states of nucleons, deltas and pions explicitly \cite{hartnack2}, as inherited from the 
Vlasov-Uehling-Uhlenbeck (VUU) model \cite{kruse}. The details about the elastic and inelastic cross sections for 
proton-proton and neutron-neutron collisions can be found in Refs.\cite{hartnack,lehmann}. \\
In this model, baryons are represented by Gaussian-shaped density distributions \\
\begin{equation}
f_i(r,p,t) = \frac{1}{{\pi}^2{\hbar}^2}e^{\frac{{-(r-r_i(t))^2}}{2L}}e^{\frac{{-(p-p_i(t))^2}.2L}{\hbar^2}}.
\end{equation}
Nucleons are initialized in a sphere with radius R = ${1.12A^{1/3}}$ fm, in accordance with the  liquid 
drop model. Each nucleon occupies a volume of ${\hbar^3}$ so that phase space is uniformly filled.
The initial momenta are randomly chosen between 0 and Fermi momentum ${p_F}$. The nucleons of the target 
and projectile interact via two and three-body Skyrme forces and Yukawa potential. The isospin degrees of 
freedom is treated explicitly by employing a symmetry potential and explicit Coulomb forces between 
protons of the colliding target and projectile. This helps in achieving the correct
distribution of protons and neutrons within the nucleus.\\
The hadrons propagate using Hamilton equations of motion:\\
\begin{equation}
\frac{d\vec{r_i}}{dt} = \frac{d<H>}{d{p_i}}~~~~;~~~~\frac{d\vec{p_i}}{dt} = -\frac{d<H>}{d{r_i}}.
\end{equation}
 with\\ 
${ <H> = <T> + <V>}$  is the Hamiltonian.
\begin{eqnarray}
   & =&  \sum_i\frac{p_i^2}{2m_i} + \sum_i \sum_{j>i}\int f_i(\vec{r},\vec{p},t)V^{ij}(\vec{r'},\vec{r})\nonumber\\ 
& &\times f_j(\vec{r'},\vec{p'},t)d\vec{r}d\vec{r'}d\vec{p}d\vec{p'}.
\end{eqnarray}
The baryon-baryon potential ${V^{ij}}$, in the above relation, reads as\\
\begin{eqnarray}
V^{ij}(\vec{r'}-\vec{r}) &~=~& V_{Skyrme}^{ij} + V_{Yukawa}^{ij} + V_{Coul}^{ij} + V_{Sym}^{ij}\nonumber\\
&=&t_1\delta(\vec{r'}-\vec{r})+ t_2\delta(\vec{r'}-\vec{r}){\rho}^{\gamma-1}(\frac{\vec{r'}+\vec{r}}{2})\nonumber\\ 
&+& t_3\frac{exp({\mid{\vec{r'}-\vec{r}}\mid}/{\mu})}{({\mid{\vec{r'}-\vec{r}}\mid}/{\mu})} + \frac{Z_{i}Z_{j}{e^2}}{\mid{\vec{r'}-\vec{r}\mid}}\nonumber\\
&+& t_{4} \frac{1}{\rho_o}T_{3}^{i}T_{3}^{j}.\delta(\vec{r'_i} - \vec{r_j}).
\end{eqnarray}
Where ${{\mu} = 0.4 fm}$, ${t_3 = -6.66 MeV}$ and ${t_4 = 100 MeV}$. 
Here ${Z_i}$ and ${Z_j}$ denote the charges of the ${i^{th}}$ and ${j^{th}}$ baryon, and ${T_{3}^i}$, 
${T_{3}^j}$ are their respective ${T_3}$ components (i.e. 1/2 for protons and -1/2 for neutrons). 
The Meson potential consists of Coulomb interaction only. The parameters ${\mu}$ and ${t_1,........,t_4}$ 
are adjusted to the real part of
the nucleonic optical potential. For the density dependence of the nucleon optical potential, standard 
Skyrme-type parameterizations is employed. The Yukawa term is quite
similar to the surface energy coefficient used in the calculations of nuclear potential for fusion
\cite{ishwar}. 
The binary nucleon-nucleon collisions are included by employing collision term of well known VUU-
Boltzmann-Uehling-Uhlenbeck (BUU) 
equation \cite{kruse, BUU}. The binary collisions are allowed stochastically, in a 
similar way as are done in 
all transport models. During the propagation, two nucleons are supposed to suffer a binary collision if 
the distance between their centroids\\
\begin{equation}
 {\mid{r_i}-{r_j}\mid} \le \sqrt{\frac{\sigma_{tot}}{\pi}},~~~~~~~ {\sigma_{tot}} = \sigma(\sqrt{s}, type),
\end{equation}
``type'' denotes the ingoing collision partners (N-N, N-${\delta}$, N-${\pi}$...). In addition, Pauli 
blocking (of the final state) of baryons is taken into account by checking the phase space densities 
in the final states. The final phase space fractions ${P_1}$ and ${P_2}$ which are already occupied by 
other nucleons,
are determined for each of the scattering baryons. The collision is then blocked with probability\\
\begin{equation}
{P_{block} = {1- (1-P_1)(1-P_2)}}.
\end{equation}  
\section{Results and Discussions}
In the present calculations, 
a simple spatial clusterization algorithm dubbed as the minimum spanning tree (MST) 
method is used to clusterize the phase space \cite{report}, which is generated by IQMD Model. We however, also acknowledge that more 
microscopic algorithm routines are also available in the literature \cite{hartnack}.
By using the asymmetric (colliding) nuclei, the effect of mass asymmetry can be analyzed without 
varying the total mass of the system. 
We have fixed (${A_{TOT}}$ = ${A_T +A_P}$ = 152) and varied the asymmetry 
of the reaction just like this: $_{26}Fe^{56}+_{44}Ru^{96}$ (${\eta = 0.2}$), $_{24}Cr^{50}+_{44}Ru^{102}$
(${\eta = 0.3}$), $_{20}Ca^{40}+_{50}Sn^{112}$ (${\eta = 0.4}$), $_{16}S^{32}+_{50}Sn^{120}$ 
(${\eta = 0.5}$), $_{14}Si^{28}+_{54}Xe^{124}$ (${\eta = 0.6}$), $_{8}O^{16}+_{54}Xe^{136}$ 
(${\eta = 0.7}$). \\
Due to the repulsive nature of Coulomb interactions, one is not able to know the exact nature of asymmetry
in the reaction dynamics.
To understand the role of asymmetry beyond the Coulomb effects, we switch off the Coulomb 
force in our analysis. Additionaly, we keep the center-of-mass energy fixed throughout the analysis.\\
\begin{figure}
\includegraphics{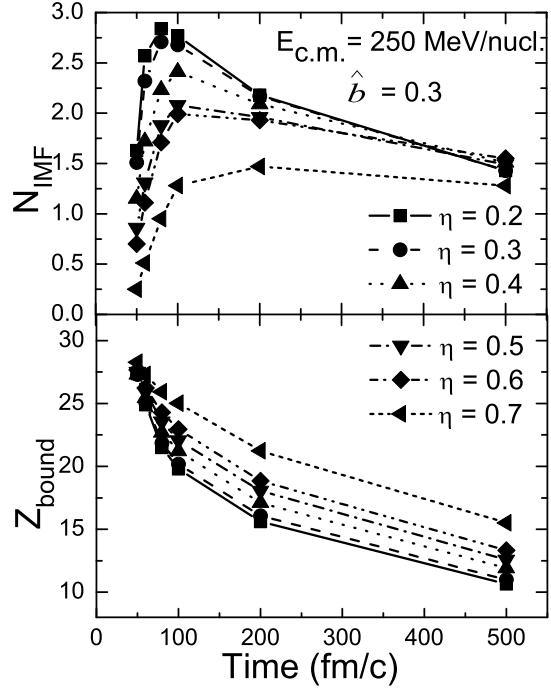}
\caption{\label{fig1} Mean multiplicity of intermediate mass fragments as a function of ${Z_{bound}}$
at ${E_{c.m.}}$ = 250 MeV/nucleon for soft equation of state. Different lines represent the different
asymmetries varying from 0.2 to 0.7. Here, the values of ${Z_{bound}}$ are recorded at different time 
steps. }
\end{figure} 
In order to study the correlation between the ${<N_{IMF}>}$
and ${Z_{bound}}$ , it is necessary to understand the time evolution of intermediate mass fragments as 
well as ${Z_{bound}}$, which is shown in Fig. 1. One learns from this figure that the mean multiplicity
of IMF increases first with the increase of time and then attains equilibrium at later times.
The system having least asymmetry gives 
rise to more IMF's as compared to system having large asymmetry. 
This might be due to the reason that as one move towards the large asymmetries, then size of the fragments
becomes larger than the size of IMFs and hence decrease in multiplicity of IMFs is observed.
On the other hand, one can see that there is a continious decrease in the value of ${Z_{bound}}$ with time 
This is due to the decay of compound nucleus into lighter particles (i.e free nucleons, LCPs etc.). It means the system is still in
non-equilibrium state. Moreover, the highly asymmetric system produces largest ${Z_{bound}}$ 
because in such a case most of the part goes uninteracted.\\ 
\begin{figure}
\includegraphics{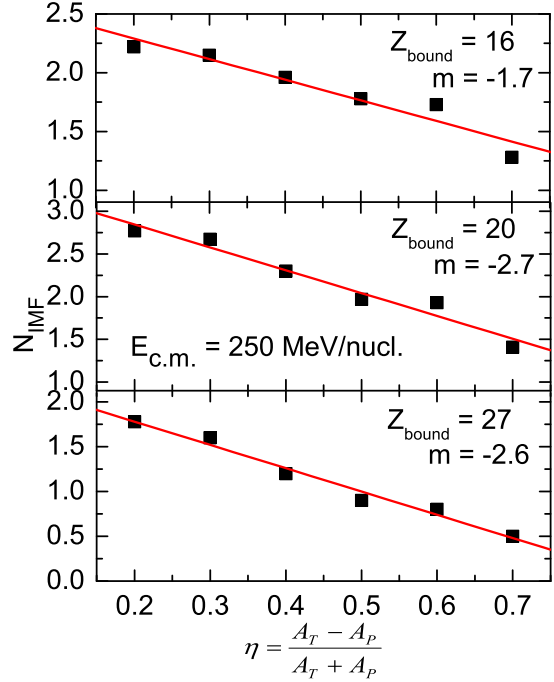}
\caption{\label{fig2} Mean multiplicity of intermediate mass fragments as a function of asymmetry
parameter ${\eta}$ at different ${Z_{bound}}$ values at ${E_{c.m.}}$ = 250 MeV/nucleon for soft equation 
of state. The lines are fitted with an equation y = mx + c, where, m represents the slope of line.  }
\end{figure} 
In Fig. 2, we show the variation of mean multiplicity of intermediate mass fragments with the asymmetry of 
the system at different values of ${Z_{bound}}$. The ${<N_{IMF}>}$ decreases with the increase in 
asymmetry 
of the system. This is true for lighter as well as heavier values of ${Z_{bound}}$. The lines are fitted 
with equation y = mx + c where, y = ${N_{IMF}}$, x = ${\eta}$, and m is slope of the line. The slope 
values are -1.7, -2.7. -2.6 corresponding to ${Z_{bound}}$
= 16, 20, 27, respectively. This indicates that for heavier ${Z_{bound}}$, production of  ${<N_{IMF}>}$ is
more sensitive with asymmetry of the system as compared to the lighter ${Z_{bound}}$ values. Moreover,
maximum IMF's are produced at ${Z_{bound}}$ = 20, indicating the limit of IMFs for a system having A = 152.\\
\begin{figure}
\includegraphics{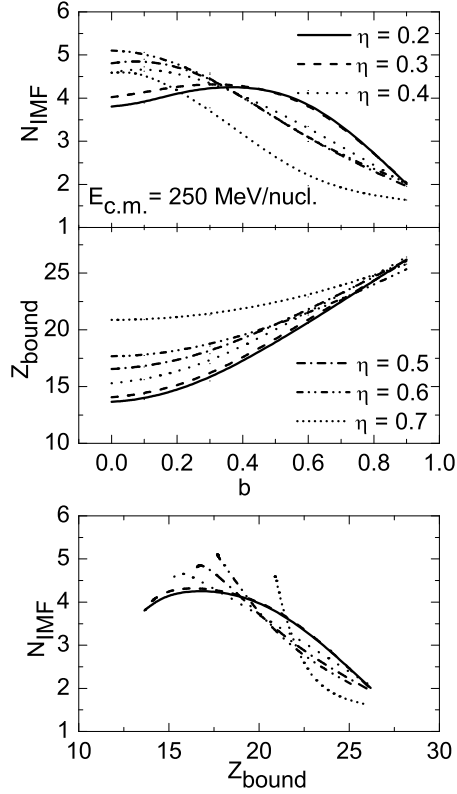}
\caption{\label{fig3} Variation of ${N_{IMF}}$ and ${Z_{bound}}$ with impact parameter at ${E_{c.m.}}$ = 
250 MeV/nucleon. Last panel shows the variation of ${N_{IMF}}$ with ${Z_{bound}}$ for different 
asymmetries varying from 0.2 to 0.7. Here, the values are recorded at different impact
parameters varying from central to peripheral one. }
\end{figure} 
Experimentalists studied many times the ${Z_{bound}}$ dependence of ${N_{IMF}}$ for symmetric 
\cite{schuttauf} as well as asymmetric systems. Following these attempts, the detailed analysis with asymmetry of the reaction is performed
in Fig. 3, where we have plotted the impact parameter dependence of ${N_{IMF}}$ (top panel), ${Z_{bound}}$ 
(medium panel), and finally ${N_{IMF}}$ versus ${Z_{bound}}$ (bottom panel).
Due to the low excitation energy E = 250 MeV/nucleon, central collisions generate repulsion in a manner 
so that the colliding nuclei breakup into IMFs, whereas for the peripheral collisions, the size of the 
fragment is close to the size of the reacting nuclei, and therefore, one sees a very few IMFs.
Interestingly, a rise and fall can be seen for 
nearly symmetric systems, which indicates the possible existence of 
various decay modes from the evaporation (fission) mode to the multifragmentation mode and then to the
vaporization mode \cite{zheng}. Moreover, this behavior is observed to be disappear with increase in
asymmetry of the reaction. On the other hand, in Fig. 3(b) the ${Z_{bound}}$ is found to 
increase with impact parameter of the reaction. As impact parameter increases, 
participant zone decreases and spectator zone increases, which will lead to the increase in the
production of heavier fragments and hence increase in the value of ${Z_{bound}}$ with impact parameter.
The symmetric systems are more sensitive with the impact parameter dependence of ${Z_{bound}}$ as
compared to the asymmetric systems. The maximum asymmetry means that from projectile or target, one is
heaviest one and other is lightest one. This leads to the possibility of IMF's production even at 
central collisions. The change in geometry does not alter too much the production of IMF's in
asymmetric systems as compared to symmetric systems. This is further elaborated in Fig. 3(c).   
which shows correlation between the ${N_{IMF}}$ and ${Z_{bound}}$ by taking into
account asymmetry of the reaction (${\eta}$ =  0.2 to 0.7). Here, the values of ${Z_{bound}}$ are 
recorded at different impact parameters varying from central to peripheral one. It was shown \cite{hubele}
that ${Z_{bound}}$ allows a very good determination of the impact parameters and hence different
reaction geometries. The smaller values of ${Z_{bound}}$ 
correspond to more central collisions. Since the IMF's are produced due to 
target breakup into pieces, therefore, as we vary the asymmetry parameter from ${\eta}$ = 0.2 to 0.7, the
target fragmentation increases. The maximum number of IMF's are observed in the ${Z_{bound}}$ 
range from 15 to 20. This is in agreement with the findings shown in Fig. 2. At lowest asymmetry, we get a 
rise and fall in the production of IMF's with ${Z_{bound}}$. But as the asymmetry increases, the curve 
shows a steep variation. These findings are supported by the findings of Fig. 3(a). 
From this, one can see that at highest asymmetry, the size of bounded fragment becomes largest. Therefore
reaction dynamics changes drastically as one moves from low asymmetry to high asymmetry.\\


\section {Conclusion}
We present our recent results on the fragmentation by varying the asymmetry of the reaction between
0.2 and 0.7 at an incident energy of 250 MeV/nucleon. For the present study, the total mass of the system 
is kept constant (${A_{TOT}}$ = 152) and asymmetry of the reaction is defined by 
(${{\eta}=  {\mid(A_T-A_P)}/{(A_T+A_P)}\mid}$). The measured distributions are given as a function of
the total charge of all projectile fragments, ${Z_{bound}}$. We see an interesting outcome for large
asymmetric colliding niclei. Although nearly symmetric nuclei depict a well known trend of rising and 
falling, this trend, however, is completely missing for large asymmetric nuclei. In conclusion, experiments
are needed to verify this prediction.   
\section {Acknowledgment}
This work has been supported by the grant from Department of Science and Technology (DST), Government of 
India, vide Grant No.SR/WOS-A/PS-10/2008.\\ 

\end{document}